\documentclass[a4paper,11pt]{article}
\pdfoutput=1
\usepackage{jheppub}

\usepackage[utf8]{inputenc}

\allowdisplaybreaks

\usepackage{amssymb}
\usepackage{amsmath}
\usepackage{breqn}
\usepackage{braket}
\usepackage{xcolor}
\usepackage{graphicx}
\usepackage{slashed}
\usepackage{bbold}

\allowdisplaybreaks

\newcommand{\e}{\epsilon}

\newcommand{\zero}{{(0)}}
\newcommand{\one}{{(1)}}
\newcommand{\two}{{(2)}}
\newcommand{\three}{{(3)}}
\newcommand{\als}{\alpha_s}
\newcommand{\alsmu}{\alpha_s(\mu)}

\newcommand{\Ord}{\mathcal{O}}

\newcommand{\nn}{\nonumber}

\newcommand{\df}{d}

\newcommand{\Gcusp}{\Gamma^{\text{cusp}}}

\def\cB{\mathcal{B}}
\def\cC{\mathcal{C}}

\def\cI{\mathcal{I}}

\def\cN{\mathcal{N}}

\DeclareMathOperator*{\SumInt}{%
\mathchoice%
  {\ooalign{$\displaystyle\sum$\cr\hidewidth$\displaystyle\int$\hidewidth\cr}}
  {\ooalign{\raisebox{.14\height}{\scalebox{.7}{$\textstyle\sum$}}\cr\hidewidth$\textstyle\int$\hidewidth\cr}}
  {\ooalign{\raisebox{.2\height}{\scalebox{.6}{$\scriptstyle\sum$}}\cr$\scriptstyle\int$\cr}}
  {\ooalign{\raisebox{.2\height}{\scalebox{.6}{$\scriptstyle\sum$}}\cr$\scriptstyle\int$\cr}}
}

\preprint{MPP-2025-160}
\title{The  N$^3$LO Twist-2 Matching of Linearly Polarized  Gluon TMDs}

\author[a]{Yu Jiao Zhu}
\emailAdd{yzhu@mpp.mpg.de}

\affiliation[a]{Max-Planck-Institut f\"{u}r Physik, Werner-Heisenberg-Institut, Boltzmannstr. 8, 85748 Garching, Germany}

\abstract{We compute the twist-2 matching  of the transverse-momentum-dependent (TMD) linearly polarized gluon parton distribution and fragmentation functions at next-to-next-to-next-to-leading order (N$^3$LO) in QCD, supplemented by next-to-next-to-leading logarithmic (NNLL) small-$x$ resummation for the gluon TMD fragmentation functions. These results provide high-precision fixed-order and resummed inputs to TMD phenomenology, and constitute essential theoretical ingredients for future studies of the spin structure and three-dimensional tomography of hadrons at the Electron-Ion Collider (EIC).
}

\begin{document}

\maketitle

\clearpage

\section{Introduction}
\label{sec:introduction}
The linearly polarized gluon distribution $h_1^{\perp g}$
contributes to observables through
quantum interference between left- and right-handed gluon states.
While perturbatively suppressed by one order in $\alpha_s$, 
the $h_1^{\perp g}$ term induces  measurable azimuthal modifications to the shape of the transverse momentum spectrum,
which has been  studied in phenomenological predictions for    diphoton processes~\cite{Nadolsky:2007ba,Boer:2013fca}, vector boson plus jet production~\cite{Dominguez:2011br,Boer:2017xpy},
dijets or heavy quark pairs productions~\cite{Boer:2009nc,Boer:2010zf,Boer:2016fqd,Efremov:2017iwh,Marquet:2017xwy,Dumitru:2015gaa,Pisano:2013cya},
and quarkonium  production~\cite{Lansberg:2017dzg,Mukherjee:2016cjw,DAlesio:2019qpk,Mukherjee:2016qxa,Lansberg:2017tlc,Kishore:2018ugo}.
The linearly polarized gluon TMDs also contribute to Higgs transverse momentum distribution~\cite{Gutierrez-Reyes:2019rug,Boer:2011kf,Echevarria:2015uaa,Catani:2010pd},
 thereby enabling precision study of   Higgs properties~\cite{Chen:2018pzu}.
In contrast, gluon TMD fragmentation function has received much less attention, despite their importance for hadronization dynamics in multi-jet events.
The three-point energy correlator in the coplanar limit with one identified hadron  can be used to access gluon TMD fragmentation dynamics~\cite{Gao:2024wcg}.

Reliable theoretical predictions within the TMD factorization framework require the perturbative matching of TMDs onto collinear PDFs at small $b_T$. 
These matching coefficients provide  boundary conditions  for both analytic resummation and global fits. 
For the unpolarized gluon distribution $f_1^g$, the matching is known up to  N$^3$LO~\cite{Ebert:2020qef,Ebert:2020yqt,Luo:2020epw}. 
For the linearly polarized gluon distribution $h_1^{\perp g}$, 
the NNLO matching coefficients were computed independently in~\cite{Gutierrez-Reyes:2019rug,Luo:2019bmw,Catani:2022sgr}.
However, the N$^3$LO matching coefficient for $h_1^{\perp g}$ has remained unknown, 
representing a key missing ingredient for extending precision predictions.

In this work, we present the first analytic computation of the N$^3$LO twist-2 matching coefficient for the linearly polarized gluon TMDs,
obtained  with an exponential regulator~\cite{Li:2016axz}.  In addition, we perform NNLL small-$x$ resummation for the coefficient functions of gluon TMD fragmentation functions. 
Together, these results provide essential theoretical input for advancing precision TMD physics.
In particular, the upcoming Electron-Ion Collider (EIC) will enable clean measurements of azimuthal asymmetries in semi-inclusive DIS, 
and dijet production,
 offering a unique opportunity to probe the transverse momentum structure and polarization of gluons inside nucleons with unprecedented accuracy. 
 Our results provide essential theoretical input for such measurements.

\section{Definition of linearly polarized gluon TMD PDFs and FFs}
\label{sec:defin-linear-gluon}
The gluon TMD PDFs can be defined in terms of SCET~\cite{Bauer:2000ew,Bauer:2000yr,Bauer:2001yt,Bauer:2002nz,Beneke:2002ph} collinear fields
\begin{align}
  \label{eq:PDFdef}
  {\cal B}_{g/N}^{{\rm bare}, \mu \nu}(x,b_\perp) =&
   - x P_+ \int \frac{d b^-}{2 \pi} e^{- i x b^- P^+} \langle N (P) | {\cal A}_{n \perp}^{a,\mu} (0, b^-, b_\perp) {\cal A}_{n \perp}^{a,\nu}(0) | N(P) \rangle \,,
\end{align}
where $N(P)$ is a hadron state with momentum $P^\mu = (\bar{n} \cdot P) n^\mu/2 = P^+ n^\mu/2$, with $n^\mu = (1, 0, 0, 1)$ and $\bar{n}^\mu = (1, 0, 0, -1)$,
and $\mathcal{A}_{n\perp}^{a,\mu}$ is the gauge invariant collinear gluon field with color index $a$ and Lorentz index $\mu$. 
Equation~(\ref{eq:PDFdef}) follows the normalization of the gluon fields in SCET  label formalism. 
An equivalent, but more conventional definition of the gluon TMDs is expressed in terms of the QCD gluon field-strength tensor with a gauge link in the adjoint representation
\begin{align}
  \label{eq:PDFdef-QCD}
   {\cal B}_{g/N}^{{\rm bare}, \mu \nu}(x,b_\perp) =&
  \frac{1}{x P_+} \int \frac{d b^-}{2 \pi} e^{- i x b^- P^+} \langle N (P) | G^{+\mu} (0, b^-, b_\perp)\mathcal{W}(0, b^-,  b_\perp;0)G^{+\nu}(0) | N(P) \rangle \,.
\end{align}
For sufficiently small $b_\perp$, the TMD PDFs in Eq.~\eqref{eq:PDFdef} admit operator product expansion onto the usual collinear PDFs~\cite{deFlorian:2001zd,Catani:2011kr,Catani:2012qa,
Gutierrez-Reyes:2017glx,Gehrmann:2012ze,Gehrmann:2014yya,
Echevarria:2016scs,Gutierrez-Reyes:2019rug,
Luo:2019hmp,Luo:2019bmw,Catani:2022sgr,Ebert:2020qef,Ebert:2020yqt,Luo:2020epw}
\begin{align}
\label{eq:PDFOPE}
 \mathcal{B}_{g/N}^{{\rm bare}, \mu \nu}(x,b_\perp) =& \sum_i \int_x^1 \frac{d \xi}{\xi} {\cal I}_{gi}^{{\rm bare}, \mu \nu} (\xi, b_\perp) \phi_{i/N}^{\rm bare}(x/\xi) + \text{power corrections}\,,
\end{align}
where the summation is over all parton flavors $i$. 
The perturbative matching coefficients $\cI^{{\rm bare}, \mu \nu}_{gi}(\xi, b_\perp)$ in Eq.~\eqref{eq:PDFOPE} are independent of the actual hadron $N$,
as a result, one can replace the hadron $N$ with a partonic state $j$ and  compute the matching coefficients within perturbation theory. 
By inserting a complete set of $n$-collinear state $\mathbb{1} = \SumInt_{X_n}  \! d {\rm PS}_{X_n}      | X_n \rangle \langle X_n |$ into the operator definition,
the bare gluon beam function can be computed from splitting amplitudes integrated over collinear phase space 
\begin{align}
\label{eq:amp-id1}
{\cal I}_{gi}^{{\rm bare}, \mu\nu} (x, b_\perp,\mu,\nu)= \lim_{\tau\to0}&\SumInt_{X_n}  \! d {\rm PS}_{X_n} 
e^{- i K_\perp \!\cdot b_\perp} 
e^{ -b_0 \tau \frac{P \cdot K}{P^+}}
\delta(K^+ - (1 - x) P^+) 
   {\rm \bold P}_{g \leftarrow i}^{\mu\nu}\,,
\end{align}
where $K^\mu$ is the total momentum of $|X_n \rangle$, and $d\text{PS}_{X_n}$ is the collinear phase space measure,
and ${\rm \bold P}_{g \leftarrow i}^{\mu\nu}$ is the spin correlator of the color averaged gluon splitting amplitude~\cite{Bern:1993qk,Bern:1994zx,Bern:1995ix,Kosower:1999xi,Bern:1999ry,Catani:1999ss,Catani:2003vu,Catani:2011st,Bern:1998sc,Kosower:1999rx,Sborlini:2013jba,
DelDuca:1999iql,Birthwright:2005ak,Birthwright:2005vi,DelDuca:2020vst,DelDuca:2019ggv,Badger:2015cxa,Czakon:2022fqi,Bern:2004cz,Badger:2004uk,Duhr:2014nda}
\begin{align}
  {\rm \bold P}_{g \leftarrow i}^{\mu\nu}
  \equiv
  {\rm \bold{Sp}}_{X_n g_{\mu}^* \leftarrow i}^{*} 
  {\rm \bold{Sp}}_{X_n g_{\nu}^* \leftarrow i} 
\equiv
- x P_+ \langle i |  {\cal A}_{n \perp}^{a,\mu}| X_n \rangle  \langle X_n |  {\cal A}_{n \perp}^{a,\nu}| i \rangle
 \,.
\end{align}
The integral in Eq.~(\ref{eq:amp-id1}) requires a rapidity cutoff to be well defined~\cite{Collins:1984kg,Ji:2004wu,Collins:2011zzd,Becher:2010tm,Becher:2011dz,Chiu:2012ir,Chiu:2009yx,Echevarria:2015byo,Li:2016axz}. In this work, 
we implement such a cutoff via the exponential rapidity regulator $e^{ -b_0 \tau \frac{P \cdot K}{P^+}}$~\cite{Li:2016axz,Luo:2019hmp},
which effectively  restricts the total energies of the collinear radiations by a rapidity scale  $\nu=1/\tau$. 
 
 A practically useful way to perform the integration in Eq.~(\ref{eq:amp-id1})
 is to first evaluate the $k_\perp$-unintegrated beam function~\cite{Luo:2019szz}
\begin{align}
\label{eq:KTbeam}
\widetilde{\mathcal{I}}&^{\mu\nu}_{g/i}(x, \widetilde K_\perp) =  \Big[ 
\lim_{\tau \to 0}  \frac{2  \int d^d K}{V_{d-2}}  e^{ -b_0 \tau \frac{P \cdot K}{P^+} } 
  \times  \delta( K^+ - P^+ (1-x) )  \delta( \widetilde K_\perp^2 - K_\perp^2 )  \nn \\
&\times\SumInt_{X_n}  \! d {\rm PS}_{X_n} \delta^{(d)}( K - \sum_{r\in{X_n}} k_r) 
 \times{\rm \bold P}_{g \leftarrow i}^{\mu\nu}(\{k_r\})
\Big]\bigg|_{\tau \to  1/\nu } \,,
\end{align}
and subsequently apply a Fourier transformation  to convert from momentum space to impact parameter space 
\begin{align}
\label{eq:kTint}
{\cal I}_{gi}^{{\rm bare}, \mu\nu} (x, b_\perp,\mu,\nu) = \int \frac{d^{d-2} \widetilde K_\perp}{|\widetilde{K}_\perp^2|^{-\e}}  e^{ - i b_\perp \cdot \widetilde K_\perp}
 \widetilde{\mathcal{I}}&^{\mu\nu}_{g/i}(x, \widetilde K_\perp)\,,
\end{align}
where $d=4 - 2 \e$ is the space-time dimension,
$b_0 = 2 e^{- \gamma_E}$ is the inverse of characteristic  transverse scale, and $V_d = 2 \pi^{d/2}/\Gamma(d/2)$ is the volume of $d$ sphere.

By boost invariance and Lorentz covariance,  the  polarization structure of the gluon beam function admit the following decomposition
\begin{align}
\label{eq:tensorStructure}
 {\cal I}_{gi}^{{\rm bare}, \mu\nu} (x, b_\perp) =&- \frac{g_\perp^{\mu\nu}}{d-2} 
{\cal I}_{gi}^{{\rm bare}}(x, b_T) + \left(-\frac{g_\perp^{\mu\nu}}{d-2} - \frac{b_\perp^\mu b_\perp^\nu}{b_T^2} \right) 
{\cal I'}_{gi}^{{\rm bare}}(x, b_T) 
\nn\\
=& -\frac{g_\perp^{\mu\nu}}{d-2}  \left[-g_{\perp{\sigma\tau}} {\cal I}_{gi}^{{\rm bare}, \sigma\tau} (x, b_\perp)\right]
\nn\\
+&  \left(-\frac{g_\perp^{\mu\nu}}{d-2} - \frac{b_\perp^\mu b_\perp^\nu}{b_T^2} \right) 
 \left[\frac{1}{d-3} \left(  -g_{\perp\sigma \tau} - \left(d-2\right) \frac{b_{\perp\sigma} b_{\perp\tau}}{b_T^2} \right) {\cal I}_{gi}^{{\rm bare}, \sigma\tau} (x, b_\perp) \right]
 \nn\\
  \widetilde{\mathcal{I}}^{\mu\nu}_{g/i}(x, \widetilde K_\perp) 
=&- \frac{g_\perp^{\mu\nu}}{d-2}  \left[ -g_{\perp{\sigma\tau}}     \widetilde{\mathcal{I}}^{\sigma\tau}_{g/i}(x, \widetilde K_\perp)      \right]
\nn\\
+&  \left(-\frac{g_\perp^{\mu\nu}}{d-2} - \frac{\widetilde K_\perp^\mu \widetilde K_\perp^\nu}{\widetilde K_T^2} \right) 
 \left[\frac{1}{d-3} \left( -g_{\perp\sigma \tau} - \left(d-2\right) \frac{\widetilde K_{\perp\sigma} \widetilde K_{\perp\tau}}{\widetilde K_T^2} \right) \widetilde{\mathcal{I}}^{\sigma\tau}_{g/i}(x, \widetilde K_\perp) \right]
\end{align} 
where $ b_T^2 = -b_\perp^2 > 0 $, $b_T = \sqrt{b_T^2} $ and $ \widetilde K_T^2 = -\widetilde K_\perp^2 > 0 $, $\widetilde K_T = \sqrt{\widetilde K_T^2} \,.$

Putting Eq.~(\ref{eq:kTint}) and Eq.~(\ref{eq:tensorStructure}) together, we arrive at
\begin{align}
{\cal I}_{gi}^{{\rm bare}, \mu\nu} (x, b_\perp) =&
-\frac{g_\perp^{\mu\nu}}{d-2} 
\int \frac{d^{d-2} \widetilde K_\perp}{|\widetilde{K}_\perp^2|^{-\e}}  e^{ - i b_\perp \cdot \widetilde K_\perp}
 \left[ -g_{\perp{\sigma\tau}}     \widetilde{\mathcal{I}}^{\sigma\tau}_{g/i}(x, \widetilde K_\perp)      \right]
 \nn\\
+&
\left(-\frac{g_\perp^{\mu\nu}}{d-2} - \frac{b_\perp^\mu b_\perp^\nu}{b_T^2} \right) 
\int \frac{d^{d-2} \widetilde K_\perp}{|\widetilde{K}_\perp^2|^{-\e}}  e^{ - i b_\perp \cdot \widetilde K_\perp}
 \bigg[
 \frac{-1+(d-2)(b_\perp\cdot \widetilde K_\perp)^2/b_T^2  \widetilde K_T^2}{(d-3)^2}
 \nn\\
 \times&
 \left( -g_{\perp\sigma \tau} - \left(d-2\right) \frac{\widetilde K_{\perp\sigma} \widetilde K_{\perp\tau}}{\widetilde K_T^2} \right) \widetilde{\mathcal{I}}^{\sigma\tau}_{g/i}(x, \widetilde K_\perp) 
\bigg]\,,
\end{align}
and thus
\begin{align}
{\cal I}_{gi}^{{\rm bare}}(x, b_T)=&
\int \frac{d^{d-2} \widetilde K_\perp}{|\widetilde{K}_\perp^2|^{-\e}}  e^{ - i b_\perp \cdot \widetilde K_\perp}
 \left[ -g_{\perp{\sigma\tau}}     \widetilde{\mathcal{I}}^{\sigma\tau}_{g/i}(x, \widetilde K_\perp)      \right]\,,
 \nn\\
{\cal I}_{gi}^{'{\rm bare}}(x, b_T)=&
\int \frac{d^{d-2} \widetilde K_\perp}{|\widetilde{K}_\perp^2|^{-\e}}  e^{ - i b_\perp \cdot \widetilde K_\perp}
 \bigg[
 \frac{-1+(d-2)(b_\perp\cdot \widetilde K_\perp)^2/b_T^2  \widetilde K_T^2}{(d-3)^2}
 \nn\\
 \times&
 \left( -g_{\perp\sigma \tau} - \left(d-2\right) \frac{\widetilde K_{\perp\sigma} \widetilde K_{\perp\tau}}{\widetilde K_T^2} \right) \widetilde{\mathcal{I}}^{\sigma\tau}_{g/i}(x, \widetilde K_\perp) 
\bigg]\,.
\end{align}
The function ${\cal I}_{gi}^{\rm bare}$ corresponds to the unpolarized gluon distribution and has been discussed extensively in our previous work~\cite{Luo:2020epw}.
 The function ${\cal I}^{'\rm bare}_{gi}$ represents the linearly polarized gluon distribution. By tensor decomposition, the  
corresponding $k_\perp$-unintegrated integrand becomes independent of the external vector $b_\perp$, 
and therefore the reduction to master integrals proceeds in  complete analogy with the unpolarized case~\cite{Luo:2019szz,Luo:2020epw}.

The TMD FFs are defined as crossings of TMD beam functions
\begin{align}
  \label{eq:FF_hadron_Frame}
  {\cal D}_{N/g}^{{\rm bare},\mu\nu} (z, b_\perp) = &
- \frac{P_+}{z^2} \sum_X \int \frac{db^-}{2 \pi} 
e^{i P^+  b^- /z} \langle 0 |
{\cal A}_{n\perp}^{a,\mu}(0, b^-, b_\perp) | N(P), X \rangle
\langle N(P), X | {\cal A}_{n\perp}^{a,\nu}(0) | 0 \rangle \,,
\end{align}
where $P^\mu = (\bar n \cdot P) n^\mu/2 =  P^+ n^\mu/2$ is the momenta of the final state detected hadron.
According to parton-hardron frame relation~\cite{Collins:2011zzd,Luo:2019hmp,Luo:2019bmw}, 
performing the calculation in the hadron frame is equivalent to working in the parton frame, 
but with the argument of the latter replaced by $b_\perp/z$ and multiplying the hadron-frame TMDs  by a flux factor $z^{2-2\epsilon}$.
For convenience, we shall henceforth denote our hadron-frame results as ${{\cal F’}}_{i/g}^{{\rm bare}}(z,b_\perp/z,\mu,\nu)$.

To obtain the finite coefficient functions, we carry out the proper UV renormalization together with the zero-bin subtraction,
which are summarized in  the following collinear mass factorization formula
\begin{align}
  \label{eq:mass-fac-form}
\frac{1}{Z_g^B}  \frac{{{\cal I}}_{g/i}^{'{\rm bare}}(x,b_\perp,\mu,\nu)}{\mathcal{S}_{0 \rm b} } = & \sum_k \mathcal{I}'_{g k}(x,b_\perp,\mu,\nu) \otimes \phi_{ki}(x,\mu)  \,,
\nn\\
\frac{1}{Z_g^B}  \frac{{{\cal F }}_{i/g}^{'{\rm bare}}(z,b_\perp/z,\mu,\nu)}{\mathcal{S}_{0 \rm b} } = & \sum_k  d_{ik}(z,\mu)\otimes \mathcal{C}'_{kg} (z, b_\perp/z,\mu,\nu)  \,.
\end{align}
where  $\mathcal{S}_{0 \rm b}(\alpha_s)$ is the bare zero-bin soft function which is the same as the TMD soft function~\cite{Li:2016ctv}, 
 $Z_g^B$ (see in Sec.~\ref{sec:RC}) are the multiplicative operator renormalization constants for the gluon correlator, and $\phi_{ki}$ ($d_{ik}$) are partonic lightcone PDFs (FFs).
$\mathcal{I}'_{g i}$ ($\mathcal{C}'_{ig}$) is the finite coefficient functions, and is one of the main results for the present work.

\section{N$^3$LO coefficients for linearly polarized  gluon TMDs}
\label{sec:n3lo-coeff-linear}
In this section we present our results for coefficient functions ${\cal I'}_{gi}$ and ${\cal C'}_{gi}$.
We will only  show the numeric fit to these functions in the paper, but the full analytic expressions can be found in the ancillary files. 
For TMD FFs, we give the results for ${\cal C}_{gi}$ with an argument $b_\perp/z$, which after divided by $z^2$ are exactly the results in the hadron frame.
\subsection{Renormalization group equations for renormalized coefficient functions }
\label{sec:RG}
The renormalized coefficient functions obey the following RG equations
\begin{align}
\frac{\df}{\df \ln\mu} \cI_{ji}^{' }(x,b_\perp,\mu,\nu) = 2 \bigg[ \Gcusp_j(\alsmu) \ln\frac{\nu}{xP_{+}} +& \gamma^B_j(\alsmu) \bigg] \cI_{ji}^{'}(x,b_\perp,\mu,\nu)
\nn\\
- &2 \sum_k \cI_{jk}^{'}(x,b_\perp,\mu,\nu) \otimes P_{ki}(x,\alsmu) \,,
\end{align}
\begin{align}
\frac{\df}{\df \ln\mu} \cC^{' }_{ij}(z,b_\perp/z,\mu,\nu) = 2 \bigg[ \Gcusp_j(\alsmu) \ln\frac{z\nu}{P_{+}} +& \gamma^B_j(\alsmu) \bigg] \cC^{'}_{ij}(z,b_\perp/z,\mu, \nu)
\nn\\
-& 2 \sum_k P^T_{ik}(z,\alsmu) \otimes \cC^{'}_{kj}(z,b_\perp/z,\mu,\nu) \, .
\label{eq:Imu}
\end{align}
The rapidity evolution equations are~\cite{Chiu:2011qc,Chiu:2012ir}
\begin{align}
\frac{\df}{\df\ln\nu} \cI_{ji}^{'}(x,b_\perp,\mu,\nu) =& -2 \left[ \int_{\mu}^{b_0/b_T} \frac{\df\bar{\mu}}{\bar{\mu}} \Gcusp_j(\alpha_s(\bar{\mu})) + \gamma^R_j(\als(b_0/b_T)) \right] \cI_{ji}^{'}(x,b_\perp,\mu,\nu) \, ,
\nn\\
\frac{\df}{\df\ln\nu} \cC^{'}_{ij}(z,b_\perp/z,\mu,\nu) =& -2 \left[ \int_{\mu}^{b_0/b_T} \frac{\df\bar{\mu}}{\bar{\mu}} \Gcusp_j(\alpha_s(\bar{\mu})) + \gamma^R_j(\als(b_0/b_T)) \right] \cC^{'}_{ij}(z,b_\perp/z,\mu, \nu) \,.
\label{eq:Inu}
\end{align}
Expanding the perturbative coefficient functions in terms of $\alpha_s/(4 \pi)$,
 the solution to these evolution equations up to $\Ord(\alpha_s^3)$ reads,
\begin{align}
\label{eq:RGS}
\cI^{'\one}_{gi}(x,b_\perp,\mu,\nu)  = & I^{'(1)}_{gi} \,, \nn
\\
\cI^{'\two}_{gi}(x,b_\perp,\mu,\nu)=&\bigg[\bigg(\beta_0-\frac{1}{2} \Gcusp_0 L_Q+\gamma_0^B\bigg)I^{'(1)}_{gi}-\sum_j I^{'(1)}_{gj}\otimes P_{ji}^\zero\bigg]L_\perp  +I^{'(2)}_{gi}\,,\nn
\\
\cI^{'\three}_{gi}(x,b_\perp,\mu,\nu)=&\bigg[\left(-\frac{3}{2}\beta_0+\frac{1}{2}\Gcusp_0 L_Q-\gamma_0^B\right) \sum_{j } I^{'(1)}_{gj}\otimes P_{ji}^\zero+\frac{1}{2}\sum_{j k}I^{'(1)}_{gj}\otimes P_{jk}^\zero\otimes P_{ki}^\zero
\nn\\
+&\frac{1}{8}
\left(
4\beta_0+2\gamma_0^B- \Gcusp_0 L_Q
\right)
\left(
2 \beta_0+2\gamma_0^B- \Gcusp_0 L_Q
\right)
I^{'(1)}_{gi}
\bigg]L_\perp^2
\nn\\
+&
\bigg[
-\sum_j I^{'(1)}_{gj}\otimes P_{ji}^\one
-\sum_j I^{'(2)}_{gj}\otimes P_{ji}^\zero
+\left(
2\beta_0-\frac{1}{2} \Gcusp_0 L_Q+\gamma_0^B
\right)
I^{'(2)}_{gi}
\nn\\
+&
\left(
\beta_1-\frac{1}{2}\Gcusp_1 L_Q+\gamma_1^B
\right)
I^{'(1)}_{gi}
\bigg]L_\perp
+I^{'(1)}_{gi} \gamma_1^R L_Q
+I^{'(3)}_{gi}
\,,
\end{align}
where  we have used $\gamma_0^R = 0$ to simplify the expression and $I^{'(n)}_{gi} $ are the scale-independent coefficient functions. 
We have defined 
\begin{align}
\label{eq:LdefinitionS}
 L_\perp = \ln \frac{b_T^2 \mu^2}{b_0^2} , \quad  L_Q  = 2 \ln \frac{x \, P_+}{\nu}, \quad L_\nu = \ln \frac{\nu^2}{\mu^2} \,,\quad b_0 =2  e^{- \gamma_E}\,.
\end{align}
The matching coefficients  of the TMD fragmentation functions allows the following form
\begin{align}
\label{eq:RGT}
\cC^{'\one}_{ig}(z,b_\perp/z,\mu,\nu)  = & C^{'(1)}_{ig} \,, \nn
\\
\cC^{'\two}_{ig}(z,b_\perp/z,\mu,\nu)=&\bigg[\bigg(\beta_0-\frac{1}{2} \Gcusp_0 L_Q+\gamma_0^B\bigg)C^{'(1)}_{ig}-\sum_j C^{'(1)}_{ij}\otimes P_{jg}^{T\zero}\bigg]L_\perp  +C^{'(2)}_{ig}\,,\nn
\\
\cC^{'\three}_{ig}(z,b_\perp/z,\mu,\nu)=&\bigg[\left(-\frac{3}{2}\beta_0+\frac{1}{2}\Gcusp_0 L_Q-\gamma_0^B\right)
 \sum_j C^{'(1)}_{ij}\otimes P_{jg}^{T\zero}+\frac{1}{2}\sum_{j k} C^{'(1)}_{ij}\otimes P_{jk}^{T\zero}\otimes P_{kg}^{T\zero}
\nn\\
+&\frac{1}{8}
\left(
4\beta_0+2\gamma_0^B- \Gcusp_0 L_Q
\right)
\left(
2 \beta_0+2\gamma_0^B- \Gcusp_0 L_Q
\right)
C^{'(1)}_{ig}
\bigg]L_\perp^2
\nn\\
+&
\bigg[
-\sum_j C^{'(1)}_{ij}\otimes P_{jg}^{T\one}
-\sum_j C^{'(2)}_{ij}\otimes P_{jg}^{T\zero}
+\left(
2\beta_0-\frac{1}{2} \Gcusp_0 L_Q+\gamma_0^B
\right)
C^{'(2)}_{ig}
\nn\\
+&
\left(
\beta_1-\frac{1}{2}\Gcusp_1 L_Q+\gamma_1^B
\right)
C^{'(1)}_{ig}
\bigg]L_\perp
+C^{'(1)}_{ig} \gamma_1^R L_Q
+C^{'(3)}_{ig}
\,,
\end{align}
We stress again that due to the chosen argument, the expressions given above are for TMD FFs in the hadron frame.
The anomalous dimensions appeared above are identical to those in the space-like case 
and the scale logarithms  are defined as
\begin{align}
\label{eq:LdefinitionT}
 L_\perp = \ln \frac{b_T^2 \mu^2}{b_0^2} , \quad L_Q = 2 \ln \frac{P_+}{ z \, \nu}, \quad L_\nu = \ln \frac{\nu^2}{\mu^2} \,,\quad b_0 =2  e^{- \gamma_E}\,,
\end{align}
Both space-like and time-like coefficient functions depend on the rapidity regulator being used. 
Rapidity-regulator-independent TMD PDFs and TMD FFs can be obtained by 
multiplying the coefficient functions with the squared root of the TMD soft functions ${\cal S}(b_\perp, \mu, \nu)$~\cite{Luo:2019hmp,Luo:2019bmw}
\begin{align}
  \label{eq:TMD-PDF-FF}
  h_{1,gi}^{\perp}(x, b_\perp, \mu) =& {\cal I}'_{gi}(x, b_\perp, \mu, \nu) \sqrt{{\cal S}(b_\perp, \mu, \nu)} \,,
\nn\\
  h_{1,ig}^{T,\perp}(z, b_\perp/z,  \mu) = &{\cal C}'_{ig}(z, b_\perp/z,  \mu, \nu) \sqrt{{\cal S}(b_\perp, \mu, \nu)} \,.
\end{align}

\subsection{Numerical fits of the N3LO coefficients}
\label{sec:num-fit}
The coefficient functions develop end-point divergences both in the threshold and high energy limit.
We first present here the results for leading threshold limit. The results for high energy limit will be discussed in next section. In the $z \to 1$ limit, we have
\begin{equation}
  \label{eq:largez}
 \lim_{z\to1}  \mathcal{I'}^{}_{gq}(z)= \lim_{z\to1} \mathcal{C'}^{}_{qg}(z) = 0\,,
\end{equation}
\begin{align}
 \lim_{z\to1}  \mathcal{I'}^{(1)}_{gg}(z)=&- \lim_{z\to1} \mathcal{C'}^{(1)}_{gg}(z) = 0\,,
\nn\\
 \lim_{z\to1}  \mathcal{I'}^{(2)}_{gg}(z)=&- \lim_{z\to1} \mathcal{C'}^{(2)}_{gg}(z) = \frac{8}{3} C_A N_f T_F-\frac{4}{3}C_A^2\,,
 \nn\\
  \lim_{z\to1}  \mathcal{I'}^{(3)}_{gg}(z)=&- \lim_{z\to1} \mathcal{C'}^{(3)}_{gg}(z) =
  C_A^3\left(-\frac{68}{9}\ln(1-z)+\frac{4}{3}\zeta_2-\frac{836}{27}\right)
  \nn\\
  +&C_A^2N_fT_F\left(\frac{152}{9}\ln(1-z)-\frac{8}{3}\zeta_2+\frac{2000}{27}\right)
  \nn\\
  +&C_A N_f^2 T_F^2\left(-\frac{32}{9}\ln(1-z)-\frac{224}{27}\right)-8 C_A C_F N_f T_F\,.
\end{align}
The analytic expressions for the  coefficient functions  will be provided in the ancillary files along with the arXiv submission.
In this section we will present their numerical fits.
 Following Ref.~\cite{Moch:2017uml}, we use the following elementary functions to fit the results,
\begin{align}
\label{eq:Lzdefinition}
L_x \equiv \ln x\,,\,L_{\bar{x}} \equiv \ln (1-x )\,, \, \bar{x}\equiv 1-x \,. 
\end{align}
We subtract the $x \to 0$ and $x \to 1$ limits up to next-to-next-to-leading power ($x^1 $ and $(1-x)^2$) and  fit the remaining terms in the region $10^{-6} < x <1$.
The fitted data in the full region $0<x<1$ has an accuracy better than $10^{-3}$.
Below we show  the numerical fitting with six significant digits for the scale independent part of the coefficient functions, the full numerical fitting is also attached as ancillary files with the arXiv submission.

\newpage
\subsubsection{Numerical fit for TMD PDFs}
The numerical fit for $q \to g$ channel reads
\begin{dmath}
I^{'(1)}_{gq}(x) =
-16.\, -5.33333 x^2+16. x+\frac{5.33333}{x}+5.33333 \bar{x}^2+5.33333 \bar{x}\,,
\end{dmath}
\begin{dmath}
I^{'(2)}_{gq}(x) =
60.0576\, +x^3 \left(1.73689 L_x^2+1.31146 L_x+181.777\right)+x^2 \left(-0.0962441 L_x^2-1.05262 L_x-485.159\right)+24.8889 L_x^2-124.444 L_x+\frac{-64. L_x-144.387}{x}+\bar{x}^3 \left(73.1746 L_{\bar{x}}-3.61139 L_{\bar{x}}^2\right)+\bar{x}^2 \left(18.3242 L_{\bar{x}}^2+109.696 L_{\bar{x}}+80.\right)+\bar{x} \left(17.7778 L_{\bar{x}}^2+39.1111 L_{\bar{x}}+167.111\right)-0.830867 x^6+1.09121 x^5-13.0824 x^4+520.649 x
+N_f \left(21.3333\, -4.58899 \bar{x}^3 L_{\bar{x}}+\bar{x}^2 \left(-6.97809 L_{\bar{x}}-4.74074\right)+\bar{x} \left(-7.11111 L_{\bar{x}}-4.74074\right)+0.0858469 x^6+0.0306905 x^5+0.857775 x^4-10.1641 x^3+29.0533 x^2-36.4561 x-\frac{4.74074}{x}\right)\,,
\end{dmath}
\begin{dmath}
I^{'(3)}_{gq}(x) =
x^3 \left(42.5798 L_x^4-193.828 L_x^3+1870.84 L_x^2-5175.95 L_x+27899.\right)+x^2 \left(-1.39715 L_x^4-46.3058 L_x^3-589.57 L_x^2-3130.51 L_x-57234.9\right)+x \left(-85.9259 L_x^3-29.037 L_x^2-496.39 L_x+33967.\right)-66.1728 L_x^4+675.951 L_x^3-5327.95 L_x^2+15498.2 L_x+\frac{384. L_x^2+6505.91 L_x+23170.5}{x}+\bar{x}^3 \left(-171.255 L_{\bar{x}}^4+427.309 L_{\bar{x}}^3-4879.47 L_{\bar{x}}^2+11122.7 L_{\bar{x}}\right)+\bar{x}^2 \left(31.6098 L_{\bar{x}}^4+472.534 L_{\bar{x}}^3+2793.57 L_{\bar{x}}^2+12374.5 L_{\bar{x}}+460.455\right)+\bar{x} \left(29.6296 L_{\bar{x}}^4+195.556 L_{\bar{x}}^3+1255.65 L_{\bar{x}}^2+3355.75 L_{\bar{x}}+4895.03\right)+5.57818 x^6+61.8984 x^5+1608.03 x^4-29477.1\, 
+N_f \left(
x^3 \left(-125.431 L_x^4+647.745 L_x^3-6056.05 L_x^2+17440.5 L_x-33441.6\right)
+5.92593 L_x^4
+x^2 \left(5.17351 L_x^4+159.374 L_x^3+2027.33 L_x^2+12536. L_x+32378.\right)
-22.1235 L_x^3+1890.26\,
+x \left(2.37037 L_x^3+33.7778 L_x^2-169.679 L_x-1858.54\right)
+\frac{-47.8311 L_x-770.159}{x} +1877.86 x^4
+\bar{x}^3 \left(26.4305 L_{\bar{x}}^3-223.37 L_{\bar{x}}^2+361.635 L_{\bar{x}}\right)
+167.506 L_x^2-245.339 L_x
+\bar{x}^2 \left(-26.6999 L_{\bar{x}}^3-219.428 L_{\bar{x}}^2-196.645 L_{\bar{x}}-492.978\right)
+11.3174 x^6-87.2009 x^5
+\bar{x} \left(-27.6543 L_{\bar{x}}^3-139.852 L_{\bar{x}}^2-419.773 L_{\bar{x}}-746.398\right)\right)
+N_f^2 \left(\bar{x}^3 \left(3.13057 L_{\bar{x}}^2+6.92382 L_{\bar{x}}\right)+\bar{x}^2 \left(7.30548 L_{\bar{x}}^2+11.6468 L_{\bar{x}}+10.2716\right)+\bar{x} \left(7.11111 L_{\bar{x}}^2+9.48148 L_{\bar{x}}+10.2716\right)-0.0798005 x^6-0.264696 x^5
-40.2963\, 
-3.35431 x^4+28.7642 x^3-65.7595 x^2+70.7187 x+\frac{10.2716}{x}\right)\,.
\end{dmath}
\newpage
The numerical fit for $g \to g$ channel reads
\begin{dmath}
I^{'(1)}_{gg}(x) =
-36.\, -12. x^2+36. x+\frac{12.}{x}+12. \bar{x}^2+12. \bar{x}
\,,
\end{dmath}
\begin{dmath}
I^{'(2)}_{gg}(x) =
 x^3 \left(59.9902 L_x^2+25.45 L_x+571.357\right)+x^2 \left(-3.90622 L_x^2-41.48 L_x-833.343\right)+72. L_x^2-228. L_x+\frac{-144. L_x-384.871}{x}+146.24 \bar{x}^3 L_{\bar{x}}+\bar{x}^2 \left(141.37 L_{\bar{x}}+154.\right)
 +292. \bar{x}+182.871\,-12.5201 x^6+43.8218 x^5-202.96 x^4+623.644 x
+N_f \left(33.3333\, -5.33333 L_x^2-8. L_x+13.3333 x^2-30.6667 x
-\frac{12.}{x}-22.6667 \bar{x}^2-21.3333 \bar{x}\right)
\,,
\end{dmath}
\begin{dmath}
I^{'(3)}_{gg}(x) =
N_f^2 \left(-15.4325\, +x^3 \left(0.720139 L_x^2-2.65876 L_x+8.87839\right)+x^2 \left(-2.39953 L_x^2-9.48416 L_x-34.4843\right)+x \left(7.11111 L_x^2+27.1111 L_x+12.532\right)+11.8519 L_x^2+41.1724 L_x+0.557351 \bar{x}^3 L_{\bar{x}}+\bar{x}^2 \left(13.0983 L_{\bar{x}}+23.8084\right)-2.66667 L_{\bar{x}}+\bar{x} \left(2.66667 L_{\bar{x}}+29.9259\right)-0.0575741 x^6+0.19494 x^5-0.96427 x^4+\frac{23.1111}{x}\right)
+N_f \left(5230.34\, +x^3 \left(-80.3575 L_x^4+304.816 L_x^3-3472.9 L_x^2+8942.15 L_x-18111.9\right)+x^2 \left(2.35664 L_x^4+73.9843 L_x^3+964.128 L_x^2+6267.36 L_x+18142.2\right)+x \left(15.7037 L_x^3+317.185 L_x^2+731.626 L_x-4550.5\right)+23.7037 L_x^4-47.7037 L_x^3
-55.9519 x^5+1290.47 x^4
+804.964 L_x^2+212.774 L_x+\frac{-94.2867 L_x-1648.92}{x}
+\bar{x} \left(97.3333 L_{\bar{x}}-1739.06\right)+1.845 x^6
+\bar{x}^3 \left(24.5475 L_{\bar{x}}^3+16.4844 L_{\bar{x}}^2+13.5722 L_{\bar{x}}\right)+\bar{x}^2 \left(3.17914 L_{\bar{x}}^3-56.1555 L_{\bar{x}}^2-340.641 L_{\bar{x}}-1231.01\right)+76. L_{\bar{x}}\right)
-75008.4\, +x^3 \left(5048.8 L_x^4-20460.3 L_x^3+220891. L_x^2-563699. L_x+1.15243\times 10^6\right)+x^2 \left(-146.656 L_x^4-4774.47 L_x^3-63544.1 L_x^2-405209. L_x-1.07734\times 10^6\right)+x \left(-360. L_x^3-1890. L_x^2-7070.08 L_x+32169.\right)-180. L_x^4+1356. L_x^3-14587.1 L_x^2+31973.9 L_x+\frac{864. L_x^2+15358.3 L_x+55590.9}{x}+\bar{x}^3 \left(893.277 L_{\bar{x}}^2+4458.05 L_{\bar{x}}\right)+\bar{x}^2 \left(602.402 L_{\bar{x}}^2+5880.86 L_{\bar{x}}+3162.92\right)-204. L_{\bar{x}}+\bar{x} \left(535.02 L_{\bar{x}}+6512.19\right)-583.156 x^6+5104.62 x^5-93142.7 x^4
\,,
\end{dmath}

\newpage
\subsubsection{Numerical fit for TMD FFs}
Similar to Eq.~(\ref{eq:Lzdefinition}), for the TMDFFs data we define 
\begin{align}
\label{eq:LzdefinitionFF}
L_z \equiv \ln z\,,\,L_{\bar{z}} \equiv \ln (1-z )\,, \, \bar{z}\equiv 1-z \,. 
\end{align}
The numerical fit for $g \to q$ channel reads
\begin{dmath}
C^{'(1)}_{qg}(z) =
2. \bar{z}^2-2. \bar{z}
\end{dmath}
\begin{dmath}
C^{'(2)}_{qg}(z) =
N_f \left(z \left(5.33333\, -2.66667 L_z\right)+\bar{z}^2 \left(0.888889\, -2.66667 L_{\bar{z}}\right)+z^2 \left(2.66667 L_z-2.66667\right)+\bar{z} \left(2.66667 L_{\bar{z}}+1.77778\right)-2.66667\right)+z^3 \left(0.218193 L_z^2-3.2449 L_z-3.90061\right)+z^2 \left(7.99543 L_z^2-54.7234 L_z+1.46858\right)+z \left(44. L_z^2+96. L_z-38.119\right)+6.66667 L_z+\bar{z}^3 \left(0.308136 L_{\bar{z}}^2-4.31322 L_{\bar{z}}\right)+\bar{z}^2 \left(6.65962 L_{\bar{z}}^2+25.2474 L_{\bar{z}}-45.7974\right)+\bar{z} \left(-6.66667 L_{\bar{z}}^2-14.6667 L_{\bar{z}}+3.1307\right)+0.0267835 z^6-0.0724324 z^5+0.596675 z^4+40.
\,,
\end{dmath}
\begin{dmath}
C^{'(3)}_{qg}(z) =
N_f \left(z^3 \left(-49.5407 L_z^4+291.12 L_z^3-2550.05 L_z^2+7675.49 L_z-14579.3\right)+z^2 \left(2.40456 L_z^4+62.6266 L_z^3+848.657 L_z^2+5730.18 L_z+14183.9\right)
+z \left(8.44444 L_z^4+20.4444 L_z^3+50.8148 L_z^2-395.578 L_z-157.261\right)
+2.8226 z^6-29.8543 z^5+646.547 z^4
+\frac{-4.74074 L_z^2-2.5679 L_z+1.06851}{z}-67.884\, 
-9.77778 L_z^3+6.88889 L_z^2+67.1852 L_z+\bar{z}^3 \left(-1.16164 L_{\bar{z}}^3+6.99385 L_{\bar{z}}^2-96.4745 L_{\bar{z}}\right)
+\bar{z}^2 \left(-10.4184 L_{\bar{z}}^3-60.3494 L_{\bar{z}}^2+77.2186 L_{\bar{z}}+227.808\right)
+\bar{z} \left(10.3704 L_{\bar{z}}^3+52.4444 L_{\bar{z}}^2+25.8201 L_{\bar{z}}-40.7011\right)\right)
+1162.44\, 
+N_f^2 \bigg(3.55556-0.352232 z^4+z^3 \left(-0.1504 L_z^2+2.15947 L_z+0.627923\right)+z^2 \left(2.66999 L_z^2-8.8481 L_z+6.38239\right)+z \left(-2.66667 L_z^2+3.55556 L_z-10.2442\right)+\bar{z}^3 \left(2.15946 L_{\bar{z}}-0.150376 L_{\bar{z}}^2\right)+\bar{z}^2 \left(2.66999 L_{\bar{z}}^2-1.73701 L_{\bar{z}}-8.47669\right)+\bar{z} \left(-2.66667 L_{\bar{z}}^2-3.55556 L_{\bar{z}}+4.92113\right)-0.0153031 z^6+0.0459082 z^5\bigg)
+z^3 \left(-45.1957 L_z^4+36.1904 L_z^3-1438.77 L_z^2+3275.74 L_z-8151.69\right)+z^2 \left(-50.0147 L_z^4+445.763 L_z^3-883.432 L_z^2+5336.81 L_z+7361.47\right)+z \left(-323.333 L_z^4-1464.15 L_z^3-3094.22 L_z^2+361.681 L_z-820.487\right)+57.8519 L_z^3-31.6667 L_z^2-229.142 L_z+\frac{32. L_z^2+10.6667 L_z-4.}{z}+\bar{z}^3 \left(18.863 L_{\bar{z}}^4-82.0146 L_{\bar{z}}^3+736.555 L_{\bar{z}}^2-1157.05 L_{\bar{z}}\right)+\bar{z}^2 \left(10.6961 L_{\bar{z}}^4+91.4333 L_{\bar{z}}^3+38.4208 L_{\bar{z}}^2-1814.23 L_{\bar{z}}-2552.26\right)+\bar{z} \left(-11.1111 L_{\bar{z}}^4-73.3333 L_{\bar{z}}^3-251.545 L_{\bar{z}}^2-52.1219 L_{\bar{z}}+747.919\right)-7.34191 z^6-1.52477 z^5+461.132 z^4
\,,
\end{dmath}

\newpage
The numerical fit for $g \to g$ channel reads
\begin{dmath}
C^{'(1)}_{gg}(z) =
12. \bar{z}-12. \bar{z}^2
\,,
\end{dmath}

\begin{dmath}
C^{'(2)}_{gg}(z) =
-42.\, +z^3 \left(-0.859465 L_z^2+119.525 L_z-75.5526\right)+z^2 \left(71.7016 L_z^2-146.862 L_z+1123.6\right)+z \left(-432. L_z^2-804. L_z-953.206\right)-144. L_z+\frac{48. L_z-4.}{z}+63.9862 \bar{z}^3 L_{\bar{z}}+\bar{z}^2 \left(-144.325 L_{\bar{z}}-298.\right)-2.21135 z^6+7.79947 z^5-42.4337 z^4+280. \bar{z}
+N_f \left(6.66667\, +z \left(16. L_z^2+24. L_z+17.3333\right)+21.3333 L_z+\frac{-7.11111 L_z-0.148148}{z}-27.8519 z^2+17.3333 \bar{z}^2-17.3333 \bar{z}\right)
\,,
\end{dmath}

\begin{dmath}
C^{'(3)}_{gg}(z) =
N_f^2 \left(-25.2286\, +z^3 \left(-18.1777 L_z^3+24.3323 L_z^2-298.898 L_z+212.883\right)+z^2 \left(-0.811839 L_z^3-21.1239 L_z^2-132.408 L_z-235.919\right)+z \left(9.48148 L_z^3-19.5556 L_z^2-36.9502 L_z-45.6169\right)-47.4074 L_z+\frac{6.32099 L_z+0.72428}{z}-11.079 \bar{z}^3 L_{\bar{z}}+\bar{z}^2 \left(-5.88457 L_{\bar{z}}-21.2899\right)+2.66667 L_{\bar{z}}+1.84883 z^6-12.3415 z^5+109.871 z^4+26.3704 \bar{z}\right)+N_f \left(-2142.97\, +z^3 \left(-210.187 L_z^4+1087.21 L_z^3-10056.2 L_z^2+29048.7 L_z-55575.\right)+z^2 \left(8.74604 L_z^4+358.925 L_z^3+2898.99 L_z^2+21674.2 L_z+52111.4\right)+z \left(-192.593 L_z^4-381.333 L_z^3-356.028 L_z^2-1500.86 L_z+2387.91\right)+149.926 L_z^3+60. L_z^2+217.119 L_z+\frac{71.1111 L_z^3+225.778 L_z^2-520.309 L_z+76.5496}{z}+\bar{z}^3 \left(-4.88908 L_{\bar{z}}^3-22.3845 L_{\bar{z}}^2-16.1436 L_{\bar{z}}\right)+\bar{z}^2 \left(-3.08125 L_{\bar{z}}^3+49.9103 L_{\bar{z}}^2+405.367 L_{\bar{z}}+998.7\right)-76. L_{\bar{z}}+\bar{z} \left(173.333 L_{\bar{z}}-1421.47\right)+10.8466 z^6-127.723 z^5+2961.39 z^4\right)+11192.3\, +z^3 \left(3639.81 L_z^4-23493.5 L_z^3+188010. L_z^2-579411. L_z+1.05992\times 10^6\right)+z^2 \left(-393.642 L_z^4-6757.5 L_z^3-58150.6 L_z^2-414850. L_z-1.07276\times 10^6\right)+z \left(3528. L_z^4+14100. L_z^3+27418.6 L_z^2+65844.1 L_z+40710.2\right)-792. L_z^3-558. L_z^2-661.554 L_z+\frac{-480. L_z^3-1464. L_z^2+2438.26 L_z-681.464}{z}+\bar{z}^3 \left(889.79 L_{\bar{z}}^2-666.606 L_{\bar{z}}\right)+\bar{z}^2 \left(-569.928 L_{\bar{z}}^2-5480.68 L_{\bar{z}}-1374.39\right)+204. L_{\bar{z}}+\bar{z} \left(331.02 L_{\bar{z}}+5651.41\right)-189.737 z^6+1817.03 z^5-39239. z^4
\,,
\end{dmath}
\subsection{Perturbative convergence}
\begin{figure}[ht!]
  \centering
  \includegraphics[width=0.8\textwidth]{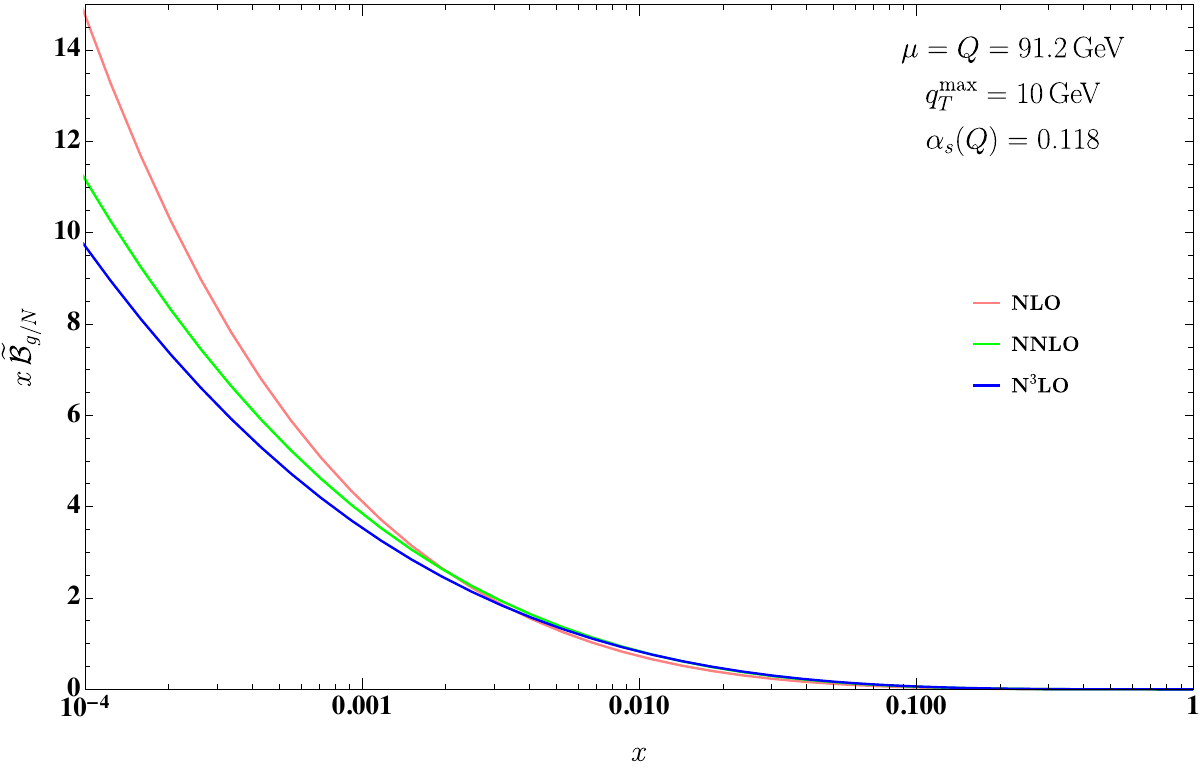}
  \caption{Integrated linearly polarized gluon TMD PDFs at various perturbative order.}
  \label{fig:beam}
\end{figure}

To investigate the perturbative convergence of (the perturbative part of)  the linearly polarized gluon TMD PDF, we consider its first moment
\begin{equation}
  \label{eq:integrated}
    h_{1,g/N}^{\perp}(x,q_T^{\rm max}) =\sum_i \int_0^{q_T^{\rm max}} \!\!\!\!\!\!\! dq_T \int_x^1 \frac{d\xi}{\xi} \,  h_{1,gi}^{\perp}(\xi, q_T, \mu) \, \phi_{i/N}(x/\xi,\mu)  \, ,
\end{equation}
where $h_{1,gi}^{\perp}(\xi, q_T, \mu)$ are physical TMD coefficients  in  momentum-space,
 and $q_T^{\rm max}$ is a UV cutoff, below which the twist-2 approximation can be justified. 
 In Fig.~(\ref{fig:beam}) we plot $x h_{1,g/N}^{\perp}$ using PDF set  \texttt{NNPDF30\_nnlo\_as\_0118}~\cite{NNPDF:2014otw}.
 We observe the perturbative uncertainties are well under control once higher-order corrections are included,
for large enough momentum fraction $x$ ($x> 10^{-3}$). 
 For extremely small $x$ ,  however, resummation effects  becomes important.
 In particular,
 the leading-logarithmic (LL) resummed prediction for the coefficient functions was obtained in  Ref.~\cite{Marzani:2015oyb}. 
%

As an application of our results, 
we consider the small transverse momentum ($p_T$)  distributions of Higgs bosons at hadron colliders,
using the frameworks of soft-collinear effective theory (SCET) \cite{Bauer:2000ew,Bauer:2000yr,Bauer:2001yt,Bauer:2002nz,Beneke:2002ph}.
At small $p_T\ll m_H=125$GeV,  
the $p_T$ distribution $d \sigma / dp_T^2$ of the produced Higgs bosons 
factorizes as an product of gluon TMD distributions and $gg\to H$ hard functions
\begin{align}
\label{eq:higgs_pt}
d \sigma / dp_T^2=&\pi \sigma_0 \int dx_a dx_b \delta(x_a x_b-\frac{m_H^2}{s})\int \frac{d^2 \vec b}{(2 \pi)^2}e^{i \vec p_T\cdot \vec b}H(m_H,\mu_F)
\nn\\
\times & S_\perp(\vec b,\mu_F,\nu) \prod_{j=a,b}\cB^{\alpha \beta}_{g/N_j}(x_j,\vec b,m_H,\mu_F,\nu)\,,
\end{align}
where  $H(m_H,\mu_F)$ is the IR-finite  function for the hard scattering  $gg\to H$ \cite{Baikov:2009bg,Lee:2010cga,Gehrmann:2010ue} and $\sigma_0$ is the corresponding born level cross section.
We  consider the cumulant of the above factorization formula by introducing a small $p_T$ cut and define the integrated distribution as 
$
\label{eq:higgs_pt_inte}
\sigma(p_T)=\int_0^{p_T} d \sigma
$,
the resulting cross section 
has a large scale logarithm $\ln(p_T^2 / m_H^2)$.  
The helicity density matrices for the unpolarized and linearly polarized  gluon distribution are orthogonal to each other,
as a result, the factorization formula in Eq.~(\ref{eq:higgs_pt}) can be schematically written as 
\begin{align}
d \sigma / dp_T^2=\frac{\pi}{2} \sigma_0 H(m_H,\mu_F) \left( h_{1,g}^{\perp} \otimes h_{1,g}^{\perp} ( p_T, \mu,m_H^2/s) 
+  f_{1,g} \otimes f_{1,g}( p_T, \mu,m_H^2/s) \right) \,.
 \end{align}
 where the first term  $ h_{1,g}^{\perp}( p_T, \mu_F)^2$  represents linearly polarized  gluon contributions,
  and the second term 
 $f_{1,g}( p_T, \mu)^2$
 corresponds to unpolarized gluon contributions.
 In what follows,  we will be particularly  concerned  with  high-order corrections to the former one.
 To this end, we calculate the contribution of linearly polarized gluons  to  the integrated small $p_T$ cross section  in Eq.~(\ref{eq:higgs_pt_inte}),
 the  renormalization and factorization scales $\mu_R$ and $\mu_F$ are chosen at typical values with $\mu_R=\mu_F=  \kappa \, m_H$ where $\kappa\in\{0.5,1,2\}$.
\begin{figure}[ht!]
  \centering
  \includegraphics[width=0.5\textwidth]{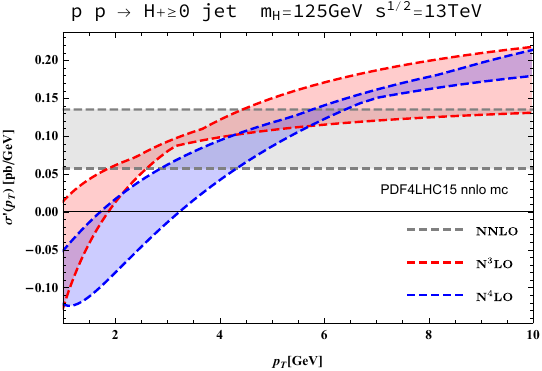}
  \caption{Linearly polarized gluon contributions to the cumulative small $p_T$ cross section, plotted as functions of the artificial cut $p_T$.
  }
  \label{fig:linear_conv}
\end{figure}
\subsection{${\cal N}=1$ supersymmetry sum rule for the linearly polarized gluon contribution}
\label{sec:cal-n=1-supersymm}
At two-loop,
it was observed that the linearly polarized gluon distribution obeys an  momentum conservation sum rule  in the ${\cal N}=1$ supersymmetric limit~\cite{Luo:2019bmw}.
Our explicit calculations confirms that this sum rule continues to hoild at three-loop.
Indeed,  by setting $C_A = C_F = N_f$ and  $T_F = 1/2$ in our perturbative bata, we find that the following sum rule is satisfied
\begin{align}
  \label{eq:lpsumrule}
  \int_0^1 dx\, x \Big( {\cal I}^{'}_{gg}(x, b_\perp, \mu,\nu) - {\cal I}^{'}_{gq}(x, b_\perp,  \mu,\nu)  \Big) \Big|_{C_F = C_A = N_f} = 0 \,.
\end{align}
To show the data explicitly, we have 
\begin{align}
  \label{eq:threeloopsumrule}
  \Big( {\cal I}^{'}_{gg}(x, b_\perp,  \mu,\nu) -& {\cal I}^{'}_{gq}(x, b_\perp,  \mu,\nu)  \Big) \Big|_{C_F = C_A = N_f} =
  \left (\frac{\alpha_s}{4 \pi} C_A\right)^2\bigg[
   -8 H_{0}-\frac{8(1-x)(x^2-2x-1)}{3x}
   \bigg]
   \nn\\
   +&
   \left (\frac{\alpha_s}{4 \pi} C_A\right)^3
   \bigg[
   L_\perp \bigg(L_Q \left (\frac{16}{3x} H_{0}+\frac{16 (x-1) (x^2-2 x-1)}{3 x} \right)
   \nn\\
   -&\frac{32 (x-1) (x^2-2 x-1) }{3 x}H_{1}-\frac{8 (4 x^3-12 x^2+19 x+4) }{3 x}H_{0}
   \nn\\
   -&32 H_{2}+16 H_{0,0}-\frac{8 (x-1) (22 x^2-17 x-65)}{9 x}+32 \zeta_2\bigg)
   \nn\\
-&\frac{8 (x-1) (5 x^2-25 x-34) }{9 x}H_{1}+\frac{16 (x+1) (x^2+2 x-2)}{3 x} H_{-1,0}
\nn\\
-&\frac{4}{3} (4 x^2-18 x+9) H_{0,0}-\frac{16 (x-1) (x^2-2 x-1) }{3 x}H_{1,0}
\nn\\
+&\left(48 \zeta_2-\frac{4 (78 x^3-72 x^2-109 x-30)}{9 x}\right) H_{0}-\frac{16 (x^3-3 x^2+7 x+4)}{3 x} H_{2}
\nn\\
-&48 H_{3}+16 H_{-2,0}-16 H_{2,0}+24 H_{0,0,0}+\frac{16 (x^3+7 x+2) }{3 x}\zeta_2
\nn\\
+&\frac{2 (x-1) (30 x^2+15 x-202)}{9 x}+40 \zeta_3
      \bigg]\,,
\end{align}
substituting this into Eq.~\eqref{eq:lpsumrule}, we indeed verify that the expression vanishes. 
which provides strong check to our three-loop results. 

\section{Small $x$ expansion of the TMD coefficients and  resummation for the TMD FFs}
\label{sec:small-x-expansion}

\subsection{Small-$x$ expansion of linearly polarized gluon TMD PDFs}
\label{sec:small-x-expansion-1}
Using the analytic expression  obtained, we can straightforwardly to obtain the small-$x$ expansion.
We find that the LL prediction  of Ref.~\cite{Marzani:2015oyb} is in agreement with our results.
To leading power in the expansion, our results are
\begin{align}
x I^{'(1)}_{gq}(x)=&C_F\,,
\nn\\
x I^{'(2)}_{gq}(x)=&C_A C_F \bigg[-16 \ln x-16 \zeta_2-\frac{88}{9}\bigg]-\frac{64}{9} C_F N_f T_F\,,
\nn\\
x I^{'(3)}_{gq}(x)=&C_A^2 C_F \bigg[\left(\frac{16}{3}\zeta_2-32 \zeta_3+\frac{15440}{27}\right) \ln x+32 \ln^2 x-432 \zeta_2-400 \zeta_3+140 \zeta_4+\frac{246713}{81}\bigg]
\nn\\
+& C_AC_F^2 \bigg[-80 \zeta_2+120 \zeta_3-216 \zeta_4+7\bigg]+C_AC_F N_f T_F \bigg[\left(\frac{256}{3}-\frac{64}{3} \zeta_2\right) \ln x+\frac{448}{9} \zeta_2
\nn\\
-&96 \zeta_3-\frac{1048}{81}\bigg]
+C_F^3 \bigg[16 \zeta_2-176 \zeta_3+256 \zeta_4+10\bigg]+C_F^2 N_f T_F \bigg[\left(\frac{128}{3} \zeta_2-\frac{6400}{27}\right) \ln x
\nn\\
+&\frac{1088}{9} \zeta_2+192 \zeta_3-\frac{10724}{9}\bigg]+\frac{832}{27} C_F N_f^2 T_F^2\,,
\end{align}
\begin{align}
x I^{'(1)}_{gg} (x)=&4 C_A\,,
\nn\\
x I^{'(2)}_{gg} (x)=&C_A^2 \bigg[-16 \ln x-16 \zeta_2-\frac{148}{9}\bigg]-\frac{136}{9} C_A N_f T_F+16 C_F N_f T_F\,,
\nn\\
x I^{'(3)}_{gg} (x)=&C_A^3 \bigg[\left(\frac{16}{3} \zeta_2-32 \zeta_3+\frac{16160}{27}\right) \ln x+32 \ln^2 x-\frac{4288}{9} \zeta_2-\frac{1432}{3} \zeta_3+180 \zeta_4+\frac{260950}{81}\bigg]
\nn\\
+&C_A^2 N_f T_F \bigg[\left(112-\frac{64}{3} \zeta_2\right) \ln x+\frac{256}{9} \zeta_2-\frac{160}{3} \zeta_3+\frac{15632}{81}\bigg]+C_F^2 N_f T_F \bigg[\frac{128}{3} \zeta_3+\frac{8}{3}\bigg]
\nn\\
+&C_A  C_F N_f T_F \bigg[\left(\frac{128}{3} \zeta_2-\frac{7840}{27}\right) \ln x+\frac{1120}{9} \zeta_2+\frac{448}{3} \zeta_3-\frac{131860}{81}\bigg]
+\frac{832}{27} C_A N_f^2 T_F^2\,.
\end{align}
\subsection{Small-$z$ expansion of linearly polarized gluon TMD FFs}
\label{sec:small-x-expansion-2}
To facilitate small-$z$ resummation for TMD FFs, we shall consider the coefficient functions in flavor singlet sector.
The flavor singlet~(denoted by a superscript$\ ^s$) coefficient functions are written in  vector form,
\begin{align}
\label{eq:singlet}
\widehat C^s(z) = 
  \begin{pmatrix}
  2 N_f C'_{qg}(z)
\\
     C'_{gg}(z)
  \end{pmatrix} \,,
\end{align}
where
$C'_{ig}(z)$ are scaleless coefficient functions as appeared in  the RG solutions \eqref{eq:RGT}.

In contrast to TMD PDFs, which generate a single logarithm at each perturbative order in the  small-$x$ limit,
TMD FFs in the singlet sector develop  double logarithms in the small-$z$ region
\begin{align}
\label{eq:double-log}
\lim_{z\to 0} z \widehat C^s_{i g}(z)= \lim_{z \to 0} z \sum_{n=1}^\infty a_s^n \widehat C^{s(n)}_{i g}(z)   { \sim}   \sum_{n=1}^\infty a_s^n \bigg(  \sum_{m=1}^{2n-2} {\ln^{2n-2-m} z} \bigg)\,,
\end{align}
where $a_s=\alpha_s/(4 \pi)$ is our perturbative expansion parameter.
The small-$z$ data in the singlet sector reads 
\begin{align}
z \widehat C^{s(1)}_{qg}(z)=&z \widehat C^{s(2)}_{qg}(z)=0\,,
\nn\\
z \widehat C^{s(3)}_{qg}(z)=&2 N_fC_A^2 T_F \bigg[\frac{64}{9} \ln^2 z+\frac{64}{27} \ln z-\frac{8}{9}\bigg]+2 N_fC_A N_f T_F^2 \bigg[-\frac{64}{27} \ln z-\frac{64}{9} \zeta_2+\frac{1040}{81}\bigg]
\nn\\
+&2 N_fC_F N_f T_F^2 \bigg[-\frac{128}{9} \ln^2 z-\frac{64}{27} \ln z+\frac{64}{9} \zeta_2-\frac{896}{81}\bigg]\,,
\nn\\
z \widehat C^{s(1)}_{gg}(z)=&0\,,
\nn\\
z \widehat C^{s(2)}_{gg}(z)=&C_A^2 \bigg[\frac{16}{3} \ln z-\frac{4}{9}\bigg]+C_F N_f T_F \bigg[\frac{16}{9}-\frac{32}{3} \ln z\bigg]-\frac{8}{9} C_A N_f T_F\,,
\nn\\
z \widehat C^{s(3)}_{gg}(z)=&C_A^3 \bigg[\left(\frac{2912}{27}-\frac{32}{3} \zeta_2\right) \ln z-\frac{160}{9} \ln^3 z-\frac{488}{9} \ln^2 z+8 \zeta_2-\frac{128}{3} \zeta_3+\frac{116}{9}\bigg]
\nn\\
+&C_A^2 N_f T_F \bigg[\left(\frac{64}{3} \zeta_2-\frac{2000}{27}\right) \ln z+\frac{32}{9} \ln^2 z+\frac{784}{9} \zeta_2+\frac{160}{3} \zeta_3-\frac{17216}{81}\bigg]
\nn\\
+& C_AC_F N_f T_F \bigg[\frac{320}{9} \ln^3 z+\frac{944}{9} \ln^2 z-\frac{560}{3} \ln z-\frac{928}{9} \zeta_2-\frac{32}{3} \zeta_3+232\bigg]
+\frac{704}{81} C_A N_f^2 T_F^2
\nn\\
+&C_F^2 N_f T_F\bigg[ 32 \ln z+\frac{128}{3} \zeta_3-\frac{152}{3}\bigg]+C_F N_f^2 T_F^2 \bigg[\frac{512}{27} \ln z-\frac{1408}{81}\bigg]\,,
\end{align}
We note that while both $\widehat C_{qg}^s$ and $\widehat C_{gg}^s$ has double logarithmic divergence in the small-$z$ limit,
 the power of leading logarithmic terms of $\widehat C_{qg}^s$ is lower by $1$ compared to the corresponding leading logarithmic terms of $\widehat C_{gg}^s$. 

\subsection{Resummation of small-$z$ logarithms for linearly polarized gluon TMD FFs }
\label{sec:resummation-small-z}
In this subsection, we  derive the all-order resummation at NNLL accuracy~(resummation of the  three highest  logarithms) in Eq.~(\ref{eq:double-log}), 
following the approach proposed in \cite{Vogt:2011jv}
and employing   the routines developed  in our previous work.
 
To this end, we start from the    collinear factorization formula in Eq.~\eqref{eq:mass-fac-form} for singlet TMD FFs~(see \eqref{eq:singlet} for the definition of singlet combination)
\begin{align}
\label{eq:massfac}
 {\cal F}_{i/g}^s(z,\e) =\frac{1}{Z_j^B}  \frac{{{\cal F }}_{i/g}^{{s,\rm bare}}(z,\e)}{\mathcal{S}_{0 \rm b} }= \sum_k  d_{ik}^s  \otimes  {C_{kg}^s(z,\e)} \,,
\end{align}
where $\mathcal{F}_{i/j}^s(z,\epsilon)$ is the unfactorized TMD fragmentation function,
 on which the usual strong coupling renormalization, zero-bin subtraction and operator renormalization have already been performed,
while the collinear mass factorization  onto the lightcone FFs has not yet been applied. 
It's also reasonable to drop out all the scale-independent terms in  Eq.~\eqref{eq:massfac},
since by RG  solutions Eq.~\eqref{eq:RGT}, they  depend on lower-order quantities and the splitting functions.

It proves convenient to work in Mellin-$N$ space
\begin{align}
  \label{eq:Mellindef}
  {\cal F}(\overline{N}, \e) = M[{\cal F}(z,\e)] :=  \int_0^1 dz \, z^{N-1} {\cal F}(z, \e)  \,,
\end{align}
where $\overline{N} = N - 1$. Small-$z$ logarithms becomes poles in $\bar N$ under Mellin transformation,
\begin{align}
M\left[\frac{1}{z}\ln^k z\right]\equiv\int_{0}^{1}dz\, z^{N-1}\frac{1}{z}\ln^k z=\frac{(-1)^k k!}{(N-1)^{k+1}}=\frac{(-1)^k k!}{\overline N^{k+1}}\,.
\end{align}
In Mellin space the  collinear factorization formula  Eq.~(\ref{eq:massfac}) becomes
\begin{align}
\label{eq:massfacN}
\left(
\begin{array}{c}
 \mathcal{F}_{qg}^s(\overline{N},\e) \\
 \mathcal{F}_{gg}^s(\overline{N},\e) \\
\end{array}
\right) = 
 {\widehat d}^s(\overline{N},\e)\cdot \left(
\begin{array}{c}
 \widehat C_{qg}^s(\overline{N},\e) \\
 \widehat C_{gg}^s(\overline{N},\e) \\
\end{array}
\right) \,,
\end{align}
where
\begin{align}
  \label{eq:dmatrix}
  \widehat d^s(\overline{N},\e) = \left(
\begin{array}{cc}
\widehat d^s_{qq}(\overline{N},\e) & \widehat d^s_{qg}(\overline{N},\e) \\
 \widehat d^s_{gq}(\overline{N},\e) & \widehat d^s_{gg}(\overline{N},\e) \\
\end{array}\right)
\end{align}
are the partonic collinear FFs in $\overline{\rm MS}$ scheme,
which evolve with the time-like splitting functions $\widehat \gamma^T(\overline{N})$
\begin{align}
\label{eq:dglap}
\frac{d }{d \ln \mu^2} \widehat d^s(\overline{N},\e)   =
2 \widehat d^s(\overline{N},\e) \cdot \widehat \gamma^T(\overline{N})\,.
\end{align}
The complete NNLO results for $\widehat \gamma^T(\overline{N})$ can be found in \cite{Chen:2020uvt}, see also \cite{Mitov:2006ic,Moch:2007tx,Almasy:2011eq}.

The crucial observation of \cite{Vogt:2011jv} is that unrenormalized collinear functions in dimensional regularization have specific asympototic behavior in the small-$z$ limit. 
In the case of TMD FFs, we can write down an general ansatz at small $z$,
\begin{align}
{\cal F}^{s(n)}_{g/g}(z,\e) =& \frac{1}{\epsilon^{2n-3}} \sum_{l=0}^{{n-2}} z^{-1-2(n-1-l) \epsilon} \big( \underbrace{c^{(1,l,n)}_{gg}}_{\text{LL}}+ \underbrace{\epsilon c^{(2,l,n)}_{gg}}_{\text{NLL}} +\underbrace{ \epsilon^2 c^{(3,l,n)}_{gg}}_{\text{NNLL}} + \dots\big) \,,
\nn\\
{\cal F}^{s(n)}_{q/g}(z,\e) =& \frac{1}{\epsilon^{2n-4}} \sum_{l=0}^{{n-3}} z^{-1-2(n-1-l) \epsilon} \big( \underbrace{c^{(1,l,n)}_{qg}}_{\text{LL}}+ \underbrace{\epsilon c^{(2,l,n)}_{qg}}_{\text{NLL}} +\underbrace{ \epsilon^2 c^{(3,l,n)}_{qg}}_{\text{NNLL}} + \dots\big) \,,
\label{eq:Fspace}
\end{align}
where $c_{gg}^{(1,l,n)}$ is the leading term  in the $\e$ expansion and small-$z$ expansion, whose knowledge correspond to LL resummation as labeled in \eqref{eq:Fspace}, and similarly for other terms. 
Precisely, for $\widehat{C}_{gg}^s(z)$ the LL series correspond to $\alpha_s^n \ln^{2 n -3}z$ terms, 
while NLL correspond to $\alpha_s^n \ln^{2n-4}z$, and NNLL to $\alpha_s^n \ln^{2n-5}z$. 
For $\widehat{C}_{qg}^s(z)$ the corresponding power of $\ln z$ is lowered by $1$.
We have verified this general ansatz through explicit N$^3$LO calculation. 
In Mellin space the corresponding ansatz reads
\begin{align}
\label{eq:ansantzNspace}
{\cal F}^{s(n)}_{g/g}(\overline{N},\e) =& \frac{1}{\epsilon^{2n-3}}  \sum_{l=0}^{{n-2}}   \frac{1}{\overline{N} -2 (n-1-l) \epsilon} \big( {c^{(1,l,n)}_{gg}}+ {\epsilon c^{(2,l,n)}_{gg}} +{ \epsilon^2 c^{(3,l,n)}_{gg}} +\dots\big)\,,
\nn\\
{\cal F}^{s(n)}_{q/g}(\overline{N},\e) =& \frac{1}{\epsilon^{2n-4}}  \sum_{l=0}^{{n-3}}  \frac{1}{\overline{N} -2 (n-1-l) \epsilon} \big( {c^{(1,l,n)}_{qg}}+ {\epsilon c^{(2,l,n)}_{qg}} +{ \epsilon^2 c^{(3,l,n)}_{qg}} +\dots\big)\,.
\end{align}
Equations  \eqref{eq:ansantzNspace} and \eqref{eq:massfacN} provides the ansatz to resum all the large logarithms of $z$. 
On the right-hand side of Eq.~(\ref{eq:massfacN}),
the $\e$ expansion begins at order $\e^{-n+1}$ for the $n$-loop contribution, 
and the non-$\e^{-1}$ infrared poles are composed of lower-order coefficient functions and splitting functions,
which can be regarded as known inputs.
On the left-hand side of Eq.~(\ref{eq:massfacN}), for example, in the unfactorized function ${\cal F}^{s(n)}_{g/g}$,
the $\e$ series starts at order $\e^{-2n+3}$ at $n$-loop as in Eq.~(\ref{eq:ansantzNspace}).
Expanding Eq.~(\ref{eq:massfacN}) up to $\e^{-n+1}$  therefore yields $n-1$ linear equations which is sufficient to determine the 
$n-1$ unknown coefficients $c^{(1,l,n)}_{gg}$ with $l=0,\cdots n-2$.
To reach NNLL accuracy,  two additional powers of $\e$-expansion are required, 
that is, we need  $\mathcal{F}^{s(n)}_{g/g}(\overline{N},\epsilon)$ up to  order $\epsilon^{-n+3}$. 
A similar analysis shows that   $\mathcal{F}^{s(n)}_{q/g}(\overline{N},\epsilon)$ must be known up to order $\epsilon^{-n+3}$ in order to achieve NNLL accuracy.
Of course, one may expand Eq.~(\ref{eq:massfacN}) further to $\e^{-1}$.
The equations generated at this order enables  the extraction of the splitting functions at the same perturbative order.
Moreover, the resulting  linear systems become over-determined, thereby providing  a non-trivial consistency check for our resummation procedure. 
 Finally,  regarding to the solution of the DGLAP evolution equation Eq.~(\ref{eq:dglap}), 
  it can be showed that $\beta_1$ or higher-order coefficients of the beta function  may safely be dropped,
  since they involve less divergent quantities in general.
 
\begin{figure}[ht!]
\centering
   \includegraphics[width=0.45\textwidth]{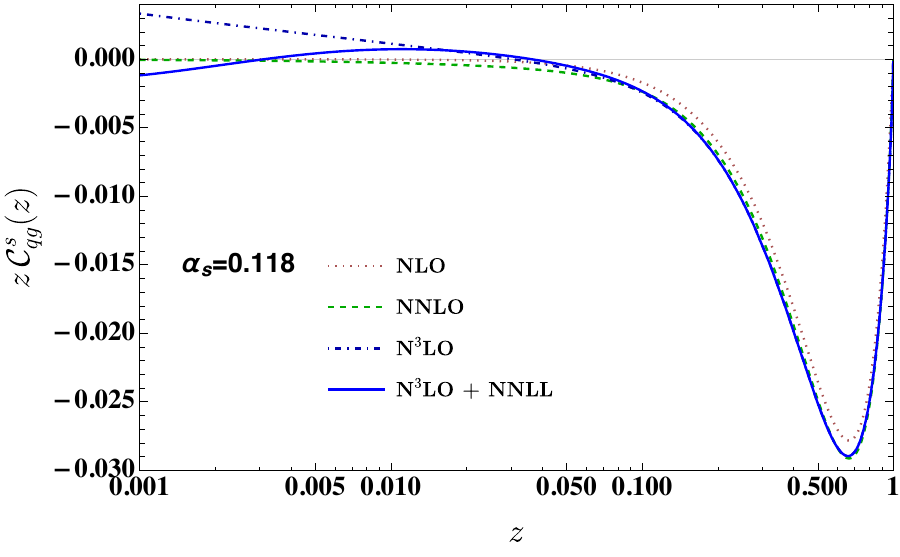}
    \includegraphics[width=0.45\textwidth]{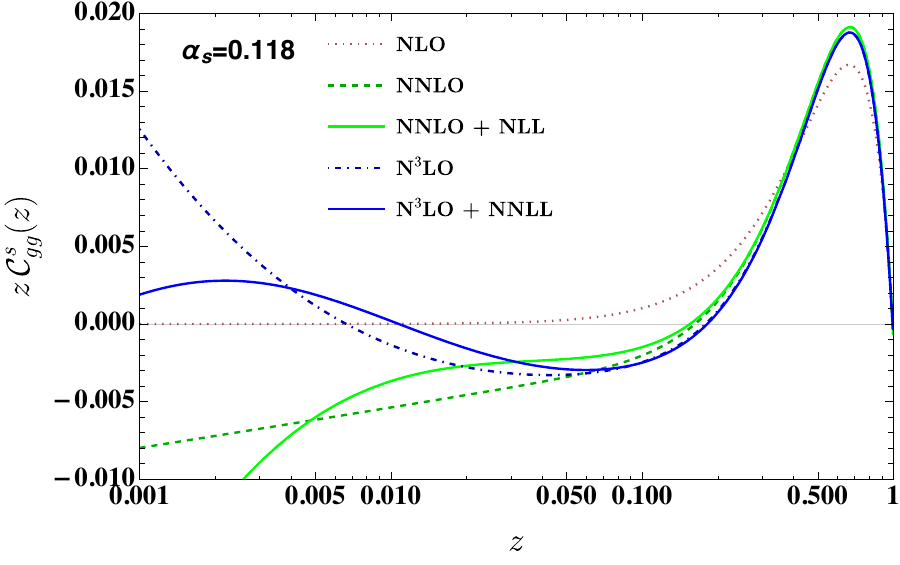}
    \caption{Coefficient functions for the linearly polarized gluon TMD FFs. 
The plots show the fixed-order results at NLO, NNLO and  N$^3$LO, 
together with the resummed predictions including
    higher-order terms truncated at order $\alpha_s^{15}$.}
\label{fig:Cg}
 \end{figure}

In summary, the input  for NNLL resummation are 
\begin{align}
&\beta_0\,,  \nonumber \\ 
& \gamma^T_0\, \,, \gamma^T_1\, \,, \gamma^T_2\, \,, \gamma^T_3\,, \nonumber\\
&{\cal F}_{g(q)/g}^{s(0)} = 0 \, \,,  \mathcal{F}_{g(q)/g}^{s(1)} \text{ to } \epsilon^2 \, \,, \mathcal{F}_{g(q)/g}^{s(2)} \text{ to }   \epsilon^1  \, \,, \mathcal{F}_{g(q)/g}^{s(3)} \text{ to }   \epsilon^0 \,.
\end{align}
Following the approach outlined above, we  obtain the resummed NNLL series truncated at order $\alpha_s^{15}$, achieving a   relative uncertainty below $0.1\%$ up to $z\sim 10^{-3}$.
The corresponding results are shown in 
 Fig.~\ref{fig:Cg}, 
 where we compare
  the fixed-order results with the resummed ones  at different orders in $\alpha_s$. 
Throughout, we set $N_f=5$ for the number of light quark flavors.
We observe that, even at N$^3$LO, the effects of resummation remain important for $z < 10^{-2}$.

\section{Conclusion}
\label{sec:conclusion}
We have presented the  N$^3$LO   twist-2 matching coefficients for linearly polarized gluon TMDs, 
together with the inclusion of next-to-next-to-leading logarithmic (NNLL) small-$x$ resummation for the coefficient functions of gluon TMD fragmentation functions.
The three-loop calculations were cross-checked in the $\cN=1$ supersymmetric limit, where the expected momentum sum rule was verified, providing a nontrivial consistency check of our results.
Our results provide essential theoretical ingredients for forthcoming experiments such as the Electron–Ion Collider, which will probe the transverse momentum structure and polarization of gluons with unprecedented accuracy, and furnish fundamental small-$x$ inputs for future studies of gluon dynamics and spin in high-energy QCD.

\acknowledgments
I thank Hua Xing Zhu for encouraging me to look into this problem and Tong-Zhi Yang for collaboration in the early stages of the project.

\appendix

\section{QCD Beta Function}
\label{sec:beta}

The QCD beta function is defined as
\begin{equation}
\frac{d\alpha_s}{d\ln\mu} = \beta(\alpha_s) = -2\alpha_s \sum_{n=0}^\infty \left( \frac{\alpha_s}{4 \pi} \right)^{n+1} \, \beta_n \, ,
\end{equation}
with~\cite{Baikov:2016tgj}
\begin{align}
\beta_0 &= \frac{11}{3} C_A - \frac{4}{3} T_F N_f \, , \nn
\\
\beta_1 &= \frac{34}{3} C_A^2 - \frac{20}{3} C_A T_F N_f - 4 C_F T_F N_f \, ,\nn
\\
\beta_2 &= \left(\frac{158 C_A}{27}+\frac{44 C_F}{9}\right) N_f^2 T_F^2 +\left(-\frac{205 C_A
   C_F}{9}-\frac{1415 C_A^2}{27}+2 C_F^2\right) N_f T_F  +\frac{2857 C_A^3}{54}\,.
\end{align}

\section{Anomalous dimension}
\label{sec:AD}

For all the anomalous dimensions entering the renormalization group equations of various TMD functions, we define the perturbative expansion in $\alpha_s$ according to
\begin{equation}
\gamma(\alpha_s) = \sum_{n=0}^\infty \left( \frac{\alpha_s}{4 \pi} \right)^{n+1} \, \gamma_n \, ,
\end{equation}

\begin{align}
\Gcusp_{0} =& 4 C_A\,, \nn
\\
\Gcusp_{1} =&  \left(\frac{268}{9}-8 
                 \zeta_2\right) C_A^2 -\frac{80 C_A T_F N_f}{9}\,, \nn
\\
\Gcusp_{2} =&\bigg[ \left(\frac{320 \zeta _2}{9}-\frac{224 \zeta _3}{3}-\frac{1672}{27}\right) C_A^2+\left(64 \zeta _3-\frac{220}{3}\right) C_F C_A\bigg] N_f T_F  \nn
   \\
 +&\left(-\frac{1072 \zeta
   _2}{9}+\frac{88 \zeta _3}{3}+88 \zeta _4+\frac{490}{3}\right) C_A^3  -\frac{64}{27} C_A
   N_f^2 T_F^2\,,\nn
     \\
\gamma^S_0 =& 0 \, , \nn
\\
\gamma^S_1 =& \left[ \left( -\frac{404}{27} + \frac{11\zeta_2}{3} + 14\zeta_3 \right) C_A   + \left( \frac{112}{27} - \frac{4 \zeta_2}{3} \right)T_F N_f   \right]  C_A   \,, \nn
\\
\gamma^S_2  =&\left(-\frac{88}{3} \zeta
   _3 \zeta _2+\frac{6325 \zeta _2}{81}+\frac{658 \zeta _3}{3}-88 \zeta
   _4-96 \zeta _5-\frac{136781}{1458}\right) C_A^3
+\left(\frac{80\zeta _2}{27}-\frac{224 \zeta _3}{27}\right.
\nn\\
+&\left.\frac{4160}{729}\right) C_A N_f^2 T_F^2\nn
    + \left(-\frac{2828 \zeta _2}{81}-\frac{728 \zeta _3}{27}+48 \zeta
   _4+\frac{11842}{729}\right) C_A^2 N_f T_F
   \nn\\
   +&\left(-4 \zeta _2-\frac{304 \zeta _3}{9}-16 \zeta
   _4+\frac{1711}{27}\right) C_F C_A N_f T_F\,.\nn
\nn\\
   \gamma^R_0 = &0 \, , \nn
\\
\gamma^R_1 = & \left[ \left( -\frac{404}{27} + 14\zeta_3 \right) C_A  +  \frac{112}{27} T_F N_f \right] C_A \,, \nn 
\\
\gamma^R_2 =&\bigg[\left(-\frac{824 \zeta _2}{81}-\frac{904 \zeta _3}{27}+\frac{20 \zeta
   _4}{3}+\frac{62626}{729}\right) C_A N_f T_F 
   +\left(-\frac{88}{3} \zeta _3 \zeta
   _2+\frac{3196 \zeta _2}{81}+\frac{6164 \zeta _3}{27}\right.
   \nn\\
   +&\left.\frac{77 \zeta _4}{3}-96 \zeta_5
   -\frac{297029}{1458}\right)C_A^2 
   + \left(-\frac{304 \zeta _3}{9}-16 \zeta_4+\frac{1711}{27}\right)  C_F N_f T_F 
   +\left(-\frac{64 \zeta_3}{9}\right.
   \nn\\
   -&\left.\frac{3712}{729}\right) N_f^2 T_F^2 \bigg] C_A \,.
\end{align}
 \begin{align}
\gamma^B_0 =& \frac{11}{3} C_A - \frac{4}{3} T_F N_f \,, \nn
\\
\gamma^B_1 =& C_A^2 \left( \frac{32}{3}+ 12 \zeta_3 \right) + \left(  -\frac{16}{3} C_A -  4 C_F \right) N_f T_F   \,, \nn
\\
\gamma^B_2 =& C_A^3\left(-80\zeta_5-16\zeta_3\zeta_2+\frac{55}{3}\zeta_4+\frac{536}{3}\zeta_3+\frac{8}{3}\zeta_2+\frac{79}{2}\right)
\nn\\
+&C_A^2 N_f T_F\left(-\frac{20}{3}\zeta_4-\frac{160}{3}\zeta_3-\frac{16}{3}\zeta_2-\frac{233}{9}\right)
+\frac{58}{9}C_A N_f^2 T_F^2
-\frac{241}{9}C_A C_F N_f T_F
\nn\\
+&2 C_F^2 N_f T_F +\frac{44}{9}C_F N_f^2 T_F^2\,.
\end{align}
The cusp anomalous dimension $\Gamma^{\text{cusp}}$ can be found in \cite{Moch:2004pa}. 
 The beam anomalous dimension $\gamma^B$ is related to the soft anomalous dimension $\gamma^S$~\cite{Li:2014afw}
  and  the hard anomalous dimensions $\gamma^H$~\cite{Moch:2005tm,Gehrmann:2010ue,Becher:2009qa} by renormalization group invariance condition $\gamma^B = \gamma^S - \gamma^H$.
 The rapidity anomalous dimension $\gamma^R$ can be found in \cite{Li:2016ctv,Vladimirov:2016dll}. 
 Note that the normalization here differ from those in \cite{Li:2016ctv} by a factor of $1/2$. 

\section{Renormalization Constants}
\label{sec:RC}
The following constants are needed for the renormalization of zero-bin subtracted~\cite{Manohar:2006nz} TMD PDFs through N$^3$LO, see e.g. Ref.~\cite{Luo:2019hmp,Luo:2019bmw}. 
The first three-order corrections to $Z^B $ and $Z^S$ are 
\begin{align}
\label{eqZqZs}
Z^B_1 =& \frac{1}{2\epsilon} \left(2 \gamma^B_0 -\Gamma_0^{\text{cusp}} L_Q \right) \,, \nonumber \\
Z^B_2 =& \frac{1}{8 \epsilon^2} \bigg( ( \Gamma_0^{\text{cusp}} L_Q - 2 \gamma^B_0)^2 + 2 \beta_0 (  \Gamma_0^{\text{cusp}} L_Q - 2 \gamma^B_0)    \bigg)
 + \frac{1}{4\epsilon} \left( 2 \gamma^B_1 - \Gamma_1^{\text{cusp}} L_Q \right) \,, \nonumber \\
Z^B_3  =& \frac{1}{48 \epsilon^3}  \left( 2 \gamma^B_0 -\Gamma^{\text{cusp}}_0 L_Q \right) \biggl( 8 \beta_0^2 + 6 \beta_0 \left( -2 \gamma^B_0 + \Gamma^{\text{cusp}}_0 L_Q \right) 
+ \left( -2 \gamma^B_0 + \Gamma^{\text{cusp}}_0 L_Q \right)^2 \biggl) \nn \\ 
+& \frac{1}{24 \epsilon^2} \biggl( \beta_1 \left(-8 \gamma^B_0 + 4  \Gamma^{\text{cusp}}_0 L_Q \right) + \left(4 \beta_0 - 6 \gamma^B_0 + 3 \Gamma^{\text{cusp}}_0 L_Q \right) \left( -2 \gamma^B_1 + \Gamma^{\text{cusp}}_1 L_Q \right)  \biggl) 
 \nn  \\    
  +& \frac{1}{6 \epsilon} \biggl(  2 \gamma^B_2 -  \Gamma^{\text{cusp}}_2 L_Q  \biggl) 
  \nn\\  
Z^S_1 =& \frac{1}{\epsilon^2} \Gamma^{\text{cusp}}_0  +  \frac{1}{\epsilon} \left( -2 \gamma^S_0 - \Gamma_0^{\text{cusp}} L_\nu \right) \,,\nonumber \\
Z^S_2 =& \frac{1}{2 \epsilon^4} (\Gamma^{\text{cusp}}_0)^2 - \frac{1}{4 \epsilon^3} \bigg(\Gamma^{\text{cusp}}_0 (3 \beta_0 + 8 \gamma^S_0)+4( \Gamma^{\text{cusp}}_0)^2 L_\nu\bigg)   - \frac{1}{2 \epsilon} \left( 2 \gamma^S_1 +  \Gamma^{\text{cusp}}_1 L_\nu \right) \nonumber \\
+& \frac{1}{4 \epsilon^2} \bigg(\Gamma^{\text{cusp}}_1 + 2 ( 2 \gamma^S_0 + \Gamma^{\text{cusp}}_0 L_\nu ) ( \beta_0 + 2 \gamma^S_0 + \Gamma^{\text{cusp}}_0 L_\nu) \bigg) \,, \nn \\
Z^S_3 =& \frac{ 1}{6 \epsilon^6} \left(\Gamma^{\text{cusp}}_0\right)^3  - \frac{1}{4 \epsilon^5} \left(\Gamma^{\text{cusp}}_0 \right)^2 \left(  3 \beta_0 + 4 \gamma^S_0 + 2 \Gamma^{\text{cusp}}_0 L_\nu \right) + \frac{1}{36 \epsilon^4 } \Gamma^{\text{cusp}}_0 \bigg( 22 \beta_0^2 + 45 \beta_0 \left(2 \gamma^S_0 + \Gamma^{\text{cusp}}_0 L_\nu \right)  \nn \\
+& 9 \left( \Gamma^{\text{cusp}}_1 + 2 \left( 2 \gamma^S_0+ \Gamma^{\text{cusp}}_0 L_\nu\right)^2  \right)   \biggl)  + \frac{1}{36 \epsilon^3} \biggl( -16 \beta_1 \Gamma^{\text{cusp}}_0 - 12 \beta_0^2 \left( 2 \gamma^S_0 + \Gamma^{\text{cusp}}_0 L_\nu \right) \nn \\
 -& 2 \beta_0 \left( 5 \Gamma^{\text{cusp}}_1 + 9 \left( 2 \gamma^S_0 + \Gamma^{\text{cusp}}_0 L_\nu \right)^2 \right) - 3 \bigg[ \Gamma^{\text{cusp}}_1 \left(6 \gamma^S_0 + 9 \Gamma^{\text{cusp}}_0 L_\nu \right)  \nn \\
+& 2 \left( 8 \left( \gamma^S_0\right)^3 + 6 \Gamma^{\text{cusp}}_0 \gamma^S_1 + 12 \Gamma^{\text{cusp}}_0  \left(\gamma^S_0\right)^2 L_\nu + 6 \left(\Gamma^{\text{cusp}}_0\right)^2 \gamma^S_0 L_\nu^2 + \left( \Gamma^{\text{cusp}}_0 \right)^3 L_\nu^3 \right) \bigg]    \biggl) \nn \\
 +& \frac{1}{18 \epsilon^2} \biggl(  2 \Gamma^{\text{cusp}}_2 + 3 \left( 2 \beta_1 \left( 2 \gamma^S_0 + \Gamma^{\text{cusp}}_0 L_\nu \right) + \left( 2 \beta_0 + 6 \gamma^S_0 + 3 \Gamma^{\text{cusp}}_0 L_\nu \right) \left( 2 \gamma^S_1 + \Gamma^{\text{cusp}}_1 L_\nu \right) \right)    \biggl) 
 \nn\\
 - & \frac{2 \gamma^S_2 + \Gamma^{\text{cusp}}_2 L_\nu}{3 \epsilon} \,.
\end{align}
We  remind the reader that the renormalization constants are formally identical for TMD PDFs and TMD FFs,
the logarithms appeared above should be replaced by their corresponding values in each case,
and we have
\begin{align}
\label{eq:LdefinitionALL}
 L_\perp = \ln \frac{b_T^2 \mu^2}{b_0^2} ,  \quad L_\nu = \ln \frac{\nu^2}{\mu^2} \,,
\end{align}
with $b_0 =2 \, e^{- \gamma_E}$ for both TMD PDFs and TMD FFs.

For  TMD PDFs,
\begin{align}
\label{eq:LdefinitionSLQ}
 L_Q  = 2 \ln \frac{x \, P_+}{\nu} \,,
\end{align}
while for TMD FFs,
\begin{align}
\label{eq:LdefinitionTLQ}
L_Q = 2 \ln \frac{P_+}{ z \, \nu}.
\end{align}

\bibliographystyle{JHEP}
\bibliography{linear_gluon}

\end{document}